\documentclass[prb,twocolumn,showpacs]{revtex4}
\usepackage{graphics}
\begin{document}
\title{Pinhole and tunneling conduction channels superimposed in magnetic 
tunnel junction: results and inferences}
\author{Soumik Mukhopadhyay}
\email{soumik.mukhopadhyay@saha.ac.in}
\author{I. Das}
\affiliation{ECMP Division, Saha Institute of Nuclear Physics, 1/AF,Bidhannagar, Kolkata 700064, India}
\begin{abstract}
The influence of ballistic channels superimposed on tunneling
conduction channels in magnetic tunnel junctions has been studied in a
manganese oxide based tunneling device.
Inversion of magnetoresistance has been observed in magnetic tunnel
junctions with pinhole nanocontacts over a broad temperature
range. The tunnel magnetoresistance undergoes a change of sign at
higher bias and temperature. This phenomenon is attributed to the
parallel conduction channels consisting of spin conserved ballistic 
transport through the pinhole contact where
the transmission probability is close to unity and spin polarized
tunneling across the insulating spacer with weak transmittivity.
The results seem to resolve a controversy regarding ballistic
magnetoresistance in ferromagnetic nanocontacts and establishes that
ballistic magnetoresistance do exist even if the previous results
are attributed to magnetostriction and magnetostatic force related artifacts. 
\end{abstract}
\maketitle
\section{Introduction}
The last decade has seen sustained efforts on part of the research community
towards achieving large room temperature magnetoresistance at low 
magnetic field via manipulation of electron spin, driven partly
due to the potential applications in memory devices and magnetic field
sensors and partly since the subject is interesting from the fundamental
point of view. Extensive studies are going on along two major
directions: $1)$ Spin polarized tunneling in ferromagnetic tunnel junction
and $2)$ Spin conserved ballistic transport across ferromagnetic 
nanocontacts. The situation can become really interesting, as will be shown
in this article, when the above two directions interfere with each other.

\indent
The story of spin polarized tunneling goes back to the early seventies
when Meservey and Tedrow~\cite{tedrow,tedrow1} 
proved that the spin of the electron tunneling from a
ferromagnet into a superconductor is conserved and that the
conductance is proportional to the density of states of the two electrodes.
These observations has led to what is now known as the Julliere 
model~\cite{julliere}
which provides a simple picture for transport in Magnetic Tunnel 
Junction (MTJ). However, it was the discovery of room temperature 
Tunnel Magnetoresistance (TMR)~\cite{jag} in $1995$ which generated 
immense interest in this field. Recently, the theoretical prediction and 
subsequent observation of large room temperature magnetoresistance
in epitaxial Fe/MgO/Fe structure~\cite{mgo} have triggered
a race for achieving even higher value of Tunnel Magnetoresistance.

\indent
Different transport regimes can be identified according to the relative 
size of various length scales corresponding to different scattering
mechanisms. An important length scale is the elastic mean free path $l$,
which estimates the average distance traversed
between successive elastic collisions with static scattering centers.
The situation $l>L$, $L$ being the typical length scale of the sample 
(usually a few nanometer), 
corresponds to the so called ballistic regime where the electron suffers 
little or no collision with static impurities and
is only limited by scattering with the boundaries of the nanosized conductor.
Study of magnetotransport properties in ferromagnetic 
nanocontacts showing conductance values of a few conductance quanta $2e^{2}/h$,
started in $1999$ with the discovery of about $300\%$ magnetoresistance
at room temperature in ballistic Ni nanocontacts~\cite{bmr1}. This
experiment was followed by several reports which claimed even higher
values of ballistic magnetoresistance~\cite{bmr2,bmr3,bmr4}.

\indent
Coming back to the experiment in ref.~\cite{bmr1}, 
initially a qualitative argument was presented in explaining the results.
It was proposed that the density of states and consequently
the spin polarization at the nanocontact is significantly larger
than bulk Ni. For sufficiently thin domain wall width the
electrons can travel across the contact in parallel configuration,
while in the antiparallel configuration, faces a strong back-scattering.
The same authors in a separate
article in the same year~\cite{bmr}, proposed that 
the essential ingredient for BMR is the condition of
nonadiabaticity in ballistic transport across the nanocontact. If the
domain wall width at the nanocontact is sufficiently thin so that the spin does
not have time to flip then the situation becomes analogous to the spin
conserved tunneling in MTJs and the magnetoresistance is related to
the transport spin polarization of the electrodes in the same way as the 
Tunneling Magnetoresistance (TMR) in Julliere~\cite{julliere} or Slonczewski's
model~\cite{slonc} which predicts positive TMR for symmetric electrode MTJ. 
It was also reported that the
ballistic magnetoresistance exhibits an universal scaling 
property~\cite{scaling}. However,
recently, this so called Ballistic Magnetoresistance (BMR) effect in 
ferromagnetic nanocontacts has attracted attention not for the
gigantic positive or negative change in resistance on application
of magnetic field but the controversy created by the report of small or
no BMR in mechanical ferromagnetic break junctions~\cite{artifact} 
and the possibility
of measurement artifact arising out of the mechanical distortion of
the contact geometry on application of magnetic field due to  
magnetostriction or magnetostatic forces~\cite{artifact,artifact1}.
It is argued that the very large changes in resistance are due to the
effect of magnetostriction or magnetostatic forces that cause the contact
to break and reform as the magnetic field is varied and that BMR does not
exist. Interestingly, 
there is another report of extremely large magnetoresistance
in oxide based ferromagnetic nanocontacts, which claims that the dominant
transport mechanism is not ballistic; 
it can either be hopping or tunneling~\cite{coey}.
We will show that BMR does exist in ferromagnetic nanocontacts although 
not in that huge magnitude as has been reported earlier.

\indent
Observation of inverse tunneling magnetoresistance
(TMR; $=\delta R/R = (R_{AP}-R_{P})/R_{P}$
where $R_{AP}$, $R_{P}$ are the junction resistances
in antiparallel and parallel magnetic configuration of the MTJs respectively.)
where the conductance
in the antiparallel magnetic configuration is higher than that in the
parallel configuration, has been instrumental in understanding some of the
important aspects of spin polarized transport in MTJs. For example, the
inverse TMR observed in experiments by De Teresa et. al.~\cite{desci} and
Sharma et. al.~\cite{sharma} have proved that the transport properties of MTJ
depend not only on the ferromagnetic metal electrodes but also on the
insulator. The effect of density of states was not apparent until de
Teresa et. al. in $1999$ showed that the
TMR depends on the specific bonding mechanism at the electrode-insulator
interface and that the choice of the insulator dictates which band
is to be selected for tunneling. They prepared two sets of ``hybrid'' MTJ's
--- Co/Al$_{2}$O$_{3}$/La$_{0.7}$Sr$0.3$MnO$_{3}$ and
Co/SrTiO$_{3}$/La$_{0.7}$Sr$0.3$MnO$_{3}$ trilayers. While
the former showed positive TMR in the entire bias range, the latter
exhibited a spectacular, so called ``Inverse TMR'' in a wide voltage range.
The Inverse TMR was attributed to the negative spin polarization
of the d band of Co which is selected for tunneling by the insulator
SrTiO$_{3}$, while the normal positive TMR in the former system is
due to the selection of positively polarized s band by Al$_{2}$O$_{3}$.
A few months before, in $1999$, M. Sharma et. al. had already published
another significant result on the effect of density of states where
composite tunnel barriers were used. An inversion of spin polarization was
observed in MTJ's with Ta$_{2}$O$_{5}$/Al$_{2}$O$_{3}$
barriers. The tunneling magnetoresistance
changes sign with applied voltage. This inversion was attributed to the
spin polarizations of the electrode/Ta$_{2}$O$_{5}$ interface and the
Al$_{2}$O$_{3}$/electrode interface being opposite.
Generally inverse TMR can occur if the sign of spin polarization of the two
electrode-insulator interfaces is opposite in the
relevant bias range. This means that while for one electrode, the majority
spin tunneling density of states (DOS) is greater than the
minority spin DOS, the conduction electrons from the other electrode should
be of minority spin character. There is another
report on inversion of TMR in asymmetric electrode Ni/NiO/Co
nanowire MTJ. Tsymbal et. al.~\cite{tsymbal1}
have shown that there is a finite probability that
resonant tunneling via localized impurity state,
which is positioned asymmetrically inside the barrier
can invert the effective spin polarization of one of the electrodes
thus leading to inverse TMR. However, for this to happen, the junction
area has to be extremely small, otherwise for larger junction area
one has to sum over all the local disorder configurations while
calculating the conductance and hence the disorder driven statistical
variations in TMR will be less probable.

\indent
Unfortunately, the influence of ballistic spin dependent transport
(due to the presence of pinhole nanocontacts which connect the two
ferromagnetic electrodes) on the magnetoresistive properties of MTJs
has not been explored substantially. This problem cannot be ignored 
since nowadays the emphasis is on fabricating low resistance MTJs using 
ultra thin insulating spacer, which increases the chances of occurrence of
pinhole shorts.
Recent simulations have shown that as much as $88\%$ of the current can
flow through the pinholes~\cite{simulation} in MTJs even though the 
bias dependence of differential conductance has positive curvature. 
It was claimed~\cite{ballistic1} that ballistic channels in MTJs 
are not magnetoresistive and the opening up of a spin-independent 
conduction channel can only reduce the TMR.
We will show that the ballistic channel in MTJs are not only magnetoresistive,
it, in fact, can cause inverse tunneling magnetoresistance.
However, the contribution due to tunneling conduction channel might not be
ignored. These two can act as parallel conducting channels.
The relative contributions from the two
conduction channels -- elastic tunneling through the insulating spacer
and ballistic spin polarized transport through the narrow pinhole shorts 
-- can change as the temperature and applied bias are varied and
magnetoresistive response can change accordingly.
\section{Experiment, Results and Discussion}
\indent
The trilayer La$_{0.67}$Sr$_{0.33}$MnO$_{3}$
 (LSMO) / Ba$_{2}$LaNbO$_{6}$ (BLNO) / LSMO
was deposited on single crystalline SrTiO$_{3}$ (100) substrate
held at a temperature $800^{0}$C and oxygen pressure $400$ mTorr,
using pulsed laser deposition. BLNO has a complex cubic perovskites 
structure and can be grown epitaxially
on single crystal perovskite substrates~\cite{blno1,blno2}. The
estimated thickness of the insulating spacer from the deposition
rate calibration of BLNO is $50\AA$. The microfabrication 
was done using photo-lithography and ion-beam milling. For further details
see ref.~\cite{soumik,soumik1}.

\begin{figure}
\resizebox{8.5cm}{6.5cm}
{\includegraphics{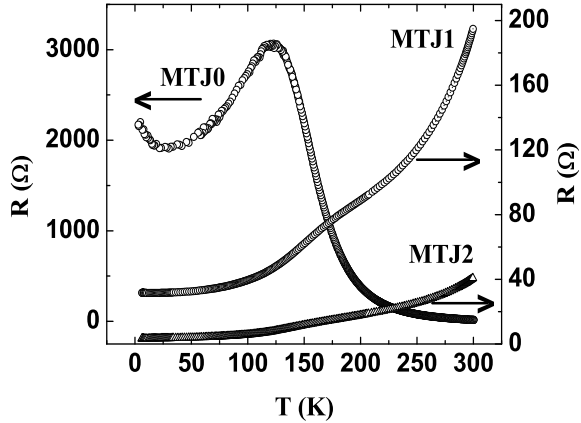}}
\caption{(A) Junction resistance vs. temperature curves for pinhole-short
MTJ1 and MTJ2 showing metal-like temperature dependence of resistance.
In contrast the MTJ without pinhole short (MTJ0) has a gross insulator-like
temperature dependence of junction resistance. The broad anomaly in
the temperature dependence of junction resistance around $130-160$ K
for MTJ1 and MTJ2 are indicative of the competition between the
tunneling and pinhole conduction channels.}\label{fig:cond}
\end{figure}

\indent
Discovery of Giaever tunneling\cite{giaever}
for superconductor-insulator-superconductor
structures in 1960's and subsequent studies gave rise to a set of criteria,
known as Rowell's criteria~\cite{rowell}, 
for determining the quality of the tunnel junction.
However, for magnetic tunnel junctions, only three of these criteria
are applicable. $1)$ Exponential thickness dependence of junction resistance.
$2)$ Parabolic differential conductance curves that should be well fitted
by rectangular barrier Simmons~\cite{simmons} 
model or trapezoidal barrier Brinkman
model~\cite{brinkman}. $3)$ Insulating like temperature
dependence of junction resistance. It has been observed that MTJs with
pinhole shorts can reproduce the first two criteria~\cite{short1}. 
Therefore the third criteria stands out as the reliable proof of the quality 
of the MTJ. The contrasting behavior in the temperature dependence of junction
resistance for three MTJs with same junction area and spacer thickness
is shown in Fig:~\ref{fig:cond}. While one of the MTJs denoted as MTJ0 
show a gross insulator like temperature dependence of junction resistance, 
the other two MTJs exhibit distinct metallic junction resistance.
\begin{figure}
\resizebox{7.5cm}{6.5cm}
{\includegraphics{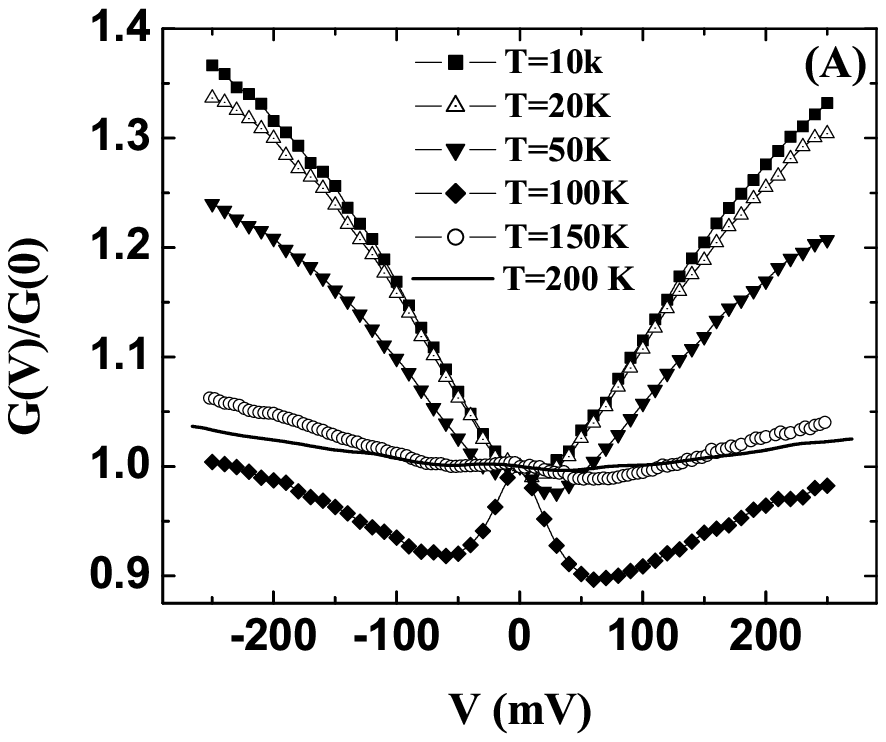}}
\resizebox{7.5cm}{6.5cm}
{\includegraphics{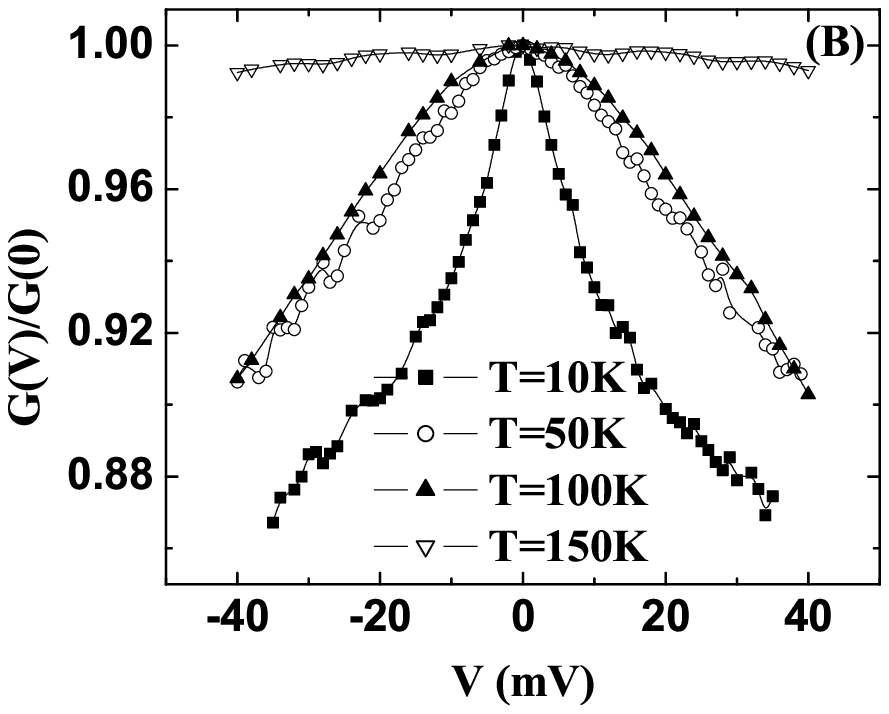}}
\caption{The voltage dependence of differential conductance for
(A) MTJ1 and (B) MTJ2 at different temperatures. While MTJ1
shows positive curvature in the voltage dependence of conductance, 
the conductance curvature for MTJ2 is negative. At higher temperatures
MTJ1 shows negative conductance curvature around zero bias,
which is an evidence of the parallel conduction channels
and the effect of opposite contributions to the sign
of curvature due to the tunneling and pinhole conduction channels.
}\label{fig:diff}
\end{figure}

\indent 
The junction resistance in the absence of magnetic field for MTJ0 
shows a distinct peak at around $125$ K with a rise at low temperature
(Fig:~\ref{fig:cond}), typical of manganite tunnel junctions~\cite{res,res1}.
Although the low temperature rise in resistance with decrease in temperature 
is consistent with Rowell's criteria for tunneling, there is an additional 
suppression of tunneling density
of states at low temperature which makes the temperature dependence
of junction resistance much sharper. Above $125$ K, the sharp decrease 
in junction resistance
with increasing temperature is attributed to the higher order tunneling
via thermally populated impurity states within the barrier.
There is still no clear explanation as to what causes the decrease in
resistance with decreasing temperature below $125$ K.
With increase in bias level the temperature dependence of junction
resistance becomes weaker. The conductance curves show parabolic voltage 
dependence. We have fitted the differential conductance vs. voltage curves
using asymmetric barrier Brinkman model~\cite{brinkman} in different voltage
ranges. The average barrier height and the barrier width turns out to be in
the range $0.2-0.3$ eV and $40\AA$ respectively.
The asymmetry in the barrier obtained from the Brinkman model is very small,
about $3-4$ mV only and hence the current-voltage characteristics can be well
fitted with symmetric barrier Simmons model~\cite{simmons}, producing similar
results. The barrier parameters like average barrier height and barrier width
are almost temperature independent within the relevant temperature range.
All these observations indicate that the device is free of any pinhole 
shorts and tunneling is the dominant transport mechanism~\cite{short}.
The highest value of TMR (as defined earlier) 
obtained at any bias current and temperature is
around $11\%$. Low TMR value signifies a considerable
reduction of spin polarization at the electrode-barrier interface. The
tunnel magnetoresistance almost vanishes above $150$ K, 
as is the case generally for manganite tunnel junctions~\cite{temp}.
\begin{figure}
\resizebox{8.5cm}{8cm}
{\includegraphics{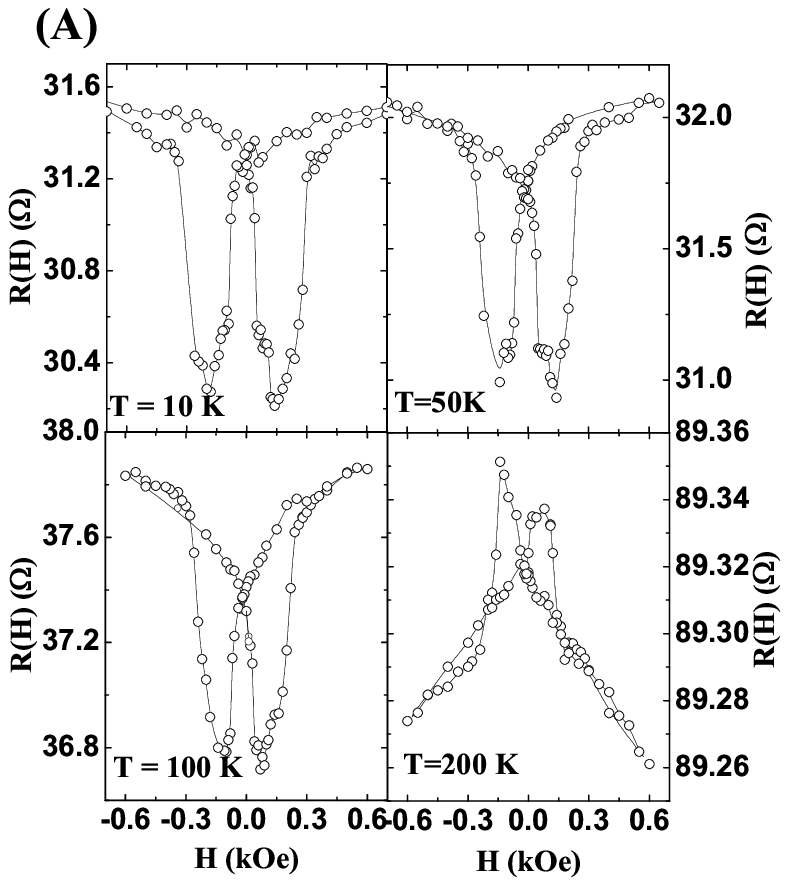}}
\resizebox{8.5cm}{8cm}
{\includegraphics{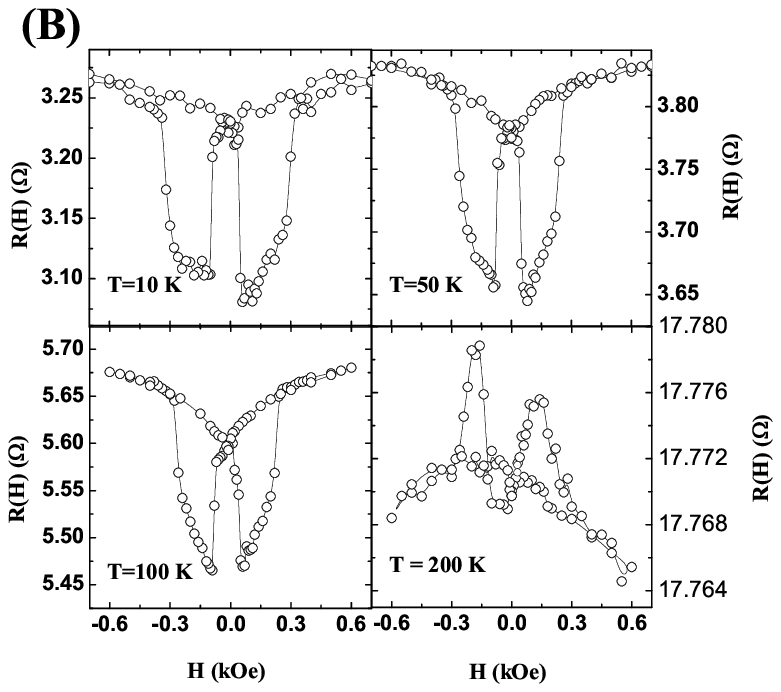}}
\caption{(A) Magnetic field dependence of junction resistance
for (A) MTJ1 and (B) MTJ2 at different temperatures showing
temperature induced inversion of magnetoresistance at $200$ K.
The value of inverse TMR is higher for MTJ2 where the contribution 
due to pinhole conduction is more compared to MTJ1.}
\label{fig:mrht1}
\end{figure}

\indent
Let us discuss the transport properties of two other MTJs (MTJ1 and MTJ2) 
fabricated under identical conditions, showing metal-like temperature 
dependence of junction resistance (Fig:~\ref{fig:cond}). Although these MTJs
show non-ohmic voltage dependence, the metallic junction resistance
is a strong evidence for the MTJs having pinhole shorts.
Each of the MTJs has the same junction area. There is orders
of magnitude difference between the junction resistance of the MTJ
without pinhole short and those having pinhole shorts. Let us designate
the pinhole shorted MTJ with higher resistance as MTJ1 and that having
lower resistance as MTJ2. The percentage rise in resistance with
increasing temperature for MTJ1 is much larger compared to that of MTJ2.
The broad anomaly in the temperature dependence of junction resistance
for MTJ1 and MTJ2 in the intermediate temperature region can be considered
as being a result of the competition between two parallel conduction
channels --- tunneling across the insulating spacer with weak transmittivity
but large effective interface area and transport through pinhole with
high transmittivity and small cross-sectional area. For MTJ1, the anomaly
shifts towards higher temperature (around $155$ K) as compared to 
MTJ2 (around $140$ K). In this article, we will show that the two MTJs with 
pinhole shorts exhibit almost identical magnetoresistive properties 
although the voltage dependence of differential conductance curves 
have opposite curvatures. While the sample denoted MTJ1 shows positive 
curvature in the conductance curve at low temperature, the conductance 
of MTJ2 has negative curvature (Fig:~\ref{fig:diff}A,B) even at $10$ K. 
The MTJs contain metallic nanocontacts through which electrons travel 
ballistically at low temperature and bias. However, at higher bias, 
``hot electron'' transport through the pinholes results in heat dissipation 
within the nanocontact region just outside the ballistic channel~\cite{heat} 
and thus increasing the resistance. At higher bias the back-scattering
into the narrow channel increases due to larger phonon density
of states at the nanocontact, which reduces the transmittivity 
resulting in negative curvature in the voltage 
dependence of differential conductance. However the conduction
channel due to tunneling will become less resistive at higher bias since then
the electrons will tunnel across relatively thin trapezoidal part
of the barrier. As a result, the pinhole short will produce 
negative curvature in the differential conductance curve while tunneling 
should cause positive curvature. Although transport in both the MTJs is 
dominated by conduction through pinhole shorts which is evident in 
Fig:~\ref{fig:cond}, the strong positive curvature in the voltage dependence 
of conductance due to tunneling can overcome the weak negative curvature due 
to transport through the pinholes, resulting in overall positive curvature as 
observed in MTJ1 (Fig:~\ref{fig:diff}A). At higher temperatures 
$50-150$ K, MTJ1 shows negative differential conductance 
around zero bias (Fig:~\ref{fig:diff}A). This anomaly can be understood 
considering that the tunneling conductance is minimum around this region 
which is evident from the temperature dependence of junction resistance for the
MTJ without pinhole short. In this temperature region the negative curvature
due to transport through pinhole short dominates at lower bias and at higher
bias positive curvature due to tunneling takes over. At $100$ K, the voltage 
at which the differential conductance takes positive curvature is higher
compared to that at $50$ and $150$ K. This is consistent with the fact
that, at $100$ K, the tunneling conductance is lower compared to that
at $50$ and $150$ K, which is evident from Fig:~\ref{fig:cond}. 
The tunneling conductance is much higher at $200$ K but still exhibits 
negative conductance curvature around zero bias, albeit small, for MTJ1 
due to the fact that at higher temperature, the positive curvature due to 
tunneling is significantly weak compared to that at low temperature.
Fitting the differential conductance curves with positive curvature for 
MTJ1 by Brinkman model, the extracted barrier height turns out to be 
about $0.8-1$ eV (much higher than the value $0.2-0.3$ eV corresponding
to MTJs without pinhole shorts) and the barrier width much smaller $15-20\AA$
compared to that of $\sim40\AA$ for good MTJs. The extracted value
for barrier height increases while the barrier width decreases as the 
temperature is increased. Although the value of the barrier parameters, 
in the present case, carry no physical significance,
temperature dependence of the barrier parameters is a reconfirmation of
the MTJ having pinhole shorts~\cite{short}. 
\begin{figure}
\resizebox{7cm}{5.5cm}
{\includegraphics{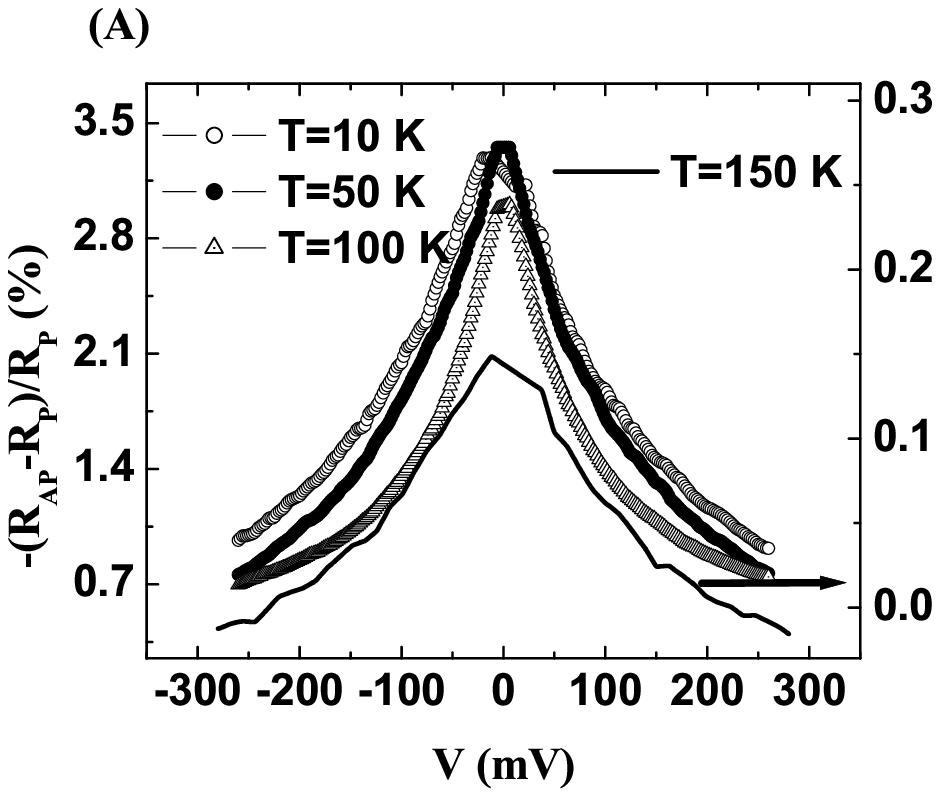}}
\resizebox{6.5cm}{5.5cm}
{\includegraphics{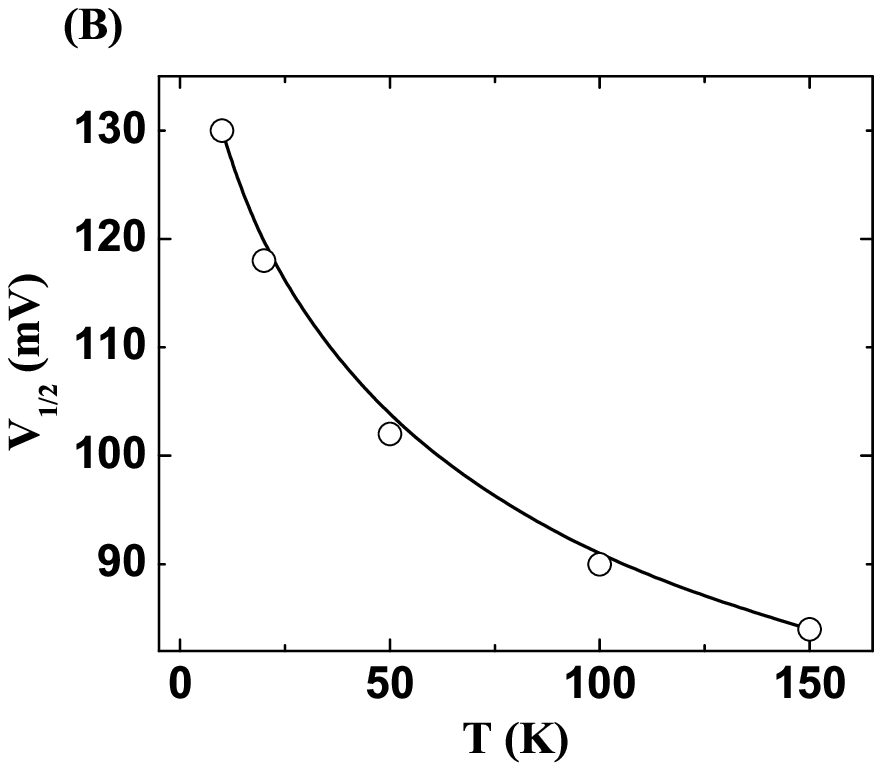}}
\caption{A: Bias dependence of TMR for MTJ1 at different temperatures
showing bias induced inversion of magnetoresistance above $\pm 225$ mV
at $150$ K. Bias dependence of TMR for MTJ1 at lower temperatures show
no evidence of sign change of TMR. B. The voltage at which TMR becomes
half its value at zero bias is plotted against corresponding temperature
value.}
\label{fig:mrbias2}
\end{figure}

\begin{figure}
\resizebox{8cm}{7cm}
{\includegraphics{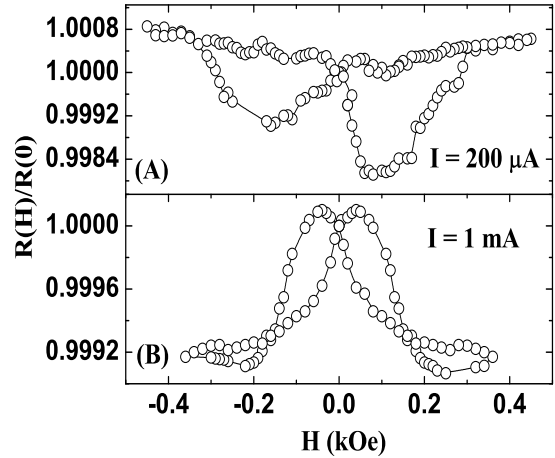}}
\caption{The reduced junction resistance
vs. magnetic field curves for MTJ1 at $150$ K at bias currents $I = 200 \mu$A
(A) and $I = 1$ mA (B). At low bias, inverse TMR is observed while at
high bias the sign of TMR reverses resulting in positive TMR.}
\label{fig:mrbias3}
\end{figure}

\begin{figure}
\resizebox{7cm}{5.5cm}
{\includegraphics{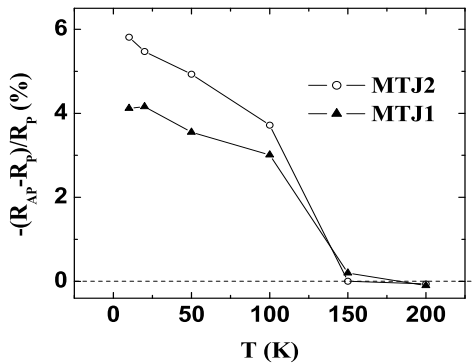}}
\caption{Temperature dependence of tunnel magnetoresistance for both
MTJ1 and MTJ2 at bias current $I=200\mu$A. There is an abrupt decrease
in the value of inverse TMR above $100$ K, indicating loss of ballisticity
at higher temperature. The TMR undergoes change of sign above $150$ K.}
\label{fig:mrbias4}
\end{figure}

\indent
Inverse TMR is observed for both MTJs over a broad temperature range 
$10-150$ K (Fig:~\ref{fig:mrht1}). The value of inverse TMR decreases 
with increasing temperature. For MTJ1 the value of inverse TMR 
is $4.6\%$ at $10$ K which reduces to about $1.8\%$ at $150$ K while for 
MTJ2 it is about $6.5\%$ at $10$ K, which almost vanishes at $150$ K.
Above $150$ K, the situation is the opposite -- ordinary positive 
TMR is observed. At $200$ K, the positive TMR exhibited 
by MTJ1 is about $0.1\%$ while that for MTJ2 is $0.06\%$ 
(Fig:~\ref{fig:mrht1}A,B). The bias dependence of TMR for MTJ1
has some interesting features. At $150$ K, it is observed that 
above $\pm 225$ mV, the TMR changes sign (Fig:~\ref{fig:mrbias2}A). 
A clear evidence of such inversion is highlighted in  
Fig:~\ref{fig:mrbias3}A,B where MTJ1 shows inverse TMR at bias 
current $I = 200 \mu A$ while at $I = 1 mA$, exhibits positive TMR. 
However, at lower temperatures,
there is no evidence of such inversion with increasing bias 
(Fig:~\ref{fig:mrbias2}A). The bias voltage at which the inverse
TMR reaches half its maximum value which is denoted by $V_{1/2}$,
becomes smaller as temperature is increased as shown in 
Fig:~\ref{fig:mrbias2}B.
It is to be noted that MTJ2 with lower junction resistance
shows higher inverse TMR values and lower positive TMR value.
The temperature dependence of TMR is described in Fig:~\ref{fig:mrbias4},
which shows an abrupt decrease in the value of inverse TMR above $100$ K.

\indent
The observed phenomenon can be explained as follows. 
The present system can be considered as being equivalent to two ferromagnetic 
metal electrodes connected by ballistic nanoscale metallic channels
along with a conduction channel connected in parallel which describes
tunneling across the insulating spacer.
For the case of two identical ferromagnets connected by a
nanocontact, the ballistic magnetoresistance (BMR)~\cite{bmr} is given by,
\[\Delta R/R_{P}=\frac{2P^{2}}{1-P^{2}}f(k_{F}\lambda)\]
where $P$ is the spin polarization, 
$\lambda$ is the domain wall width and $k_{F}$ is the Fermi wave
vector, $f$ being the measure of the spin non-conservation in the current
through the nanocontact. Because of the function $f$, the 
magnetoresistance decays rapidly for
$k_{F}\lambda \geq 1$, which indicates that the electron spin can follow
the magnetization change inside the domain wall adiabatically.
Therefore, the essential ingredient for BMR is the condition of
nonadiabaticity in ballistic transport across the nanocontact.
In the limit of vanishing domain wall width $\lambda$,
spin flipping by domain wall scattering is absent. Then $f$ is unity
and the electron spin is conserved during transmission (the factor $f$ 
decreases with the increase of the product $k_{F}\lambda$). Hence we
arrive at the well known Julliere formula for tunneling magnetoresistance.
Thus there seems to be no difference in the spin conserved ballistic transport
through nano-sized pinholes or elastic spin polarized tunneling.
However there is a stark contrast in the transmittivities for the two
conduction channels. In case of normal elastic tunneling through insulating 
barrier the tunneling probability is finite
but small and decays exponentially with increasing barrier width. 
On the other hand, the transmittivity through the metallic pinhole 
nanocontact is close to unity.
\begin{figure}
\resizebox{8cm}{6.5cm}
{\includegraphics{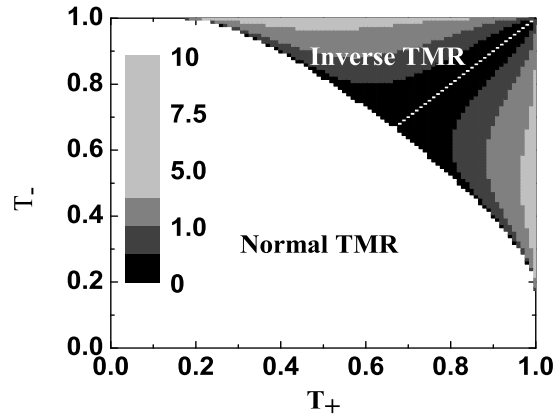}}
\caption{The theoretically allowed values of $T_{\pm}$ for
inverse TMR in the $\{T_{+},T_{-}\}$ plane and the corresponding values
of inverse TMR is shown by a shaded map. The allowed values of inverse TMR
has an upper bound of $10\%$.}
\label{fig:phase1}
\end{figure}

\indent
At low temperature, the electron transfer from one ferromagnetic lead to 
another occurs dominantly through the metallic pinhole shorts. Hence, 
according to ref.~\cite{bmr}, the ballistic magnetoresistance should follow
the Julliere or Slonczewski's model for spin polarized 
tunneling and should give positive
TMR for MTJs with identical electrodes. However, in our case, inverse
TMR is observed. The reason probably lies in the fact that the model
does not take into account the effect of high transmittivity and the 
possibility of different transmission coefficients
of the electrons in the majority and minority spin bands.
The model reduces to Julliere model in the non-adiabatic limit. 
However, there is an agreement that the generalized Julliere model
is valid in the limit of very weak transmission probability~\cite{zhang}.
Tae-Suk Kim~\cite{kim} has very recently put forward 
a theoretical model for spin polarized transport
through a narrow channel. Using the transfer Hamiltonian approach and
the non-equilibrium Green's function method, Kim has shown that
when the spin is conserved in transport through a nanoscale channel
and the transmittivity is close to unity, there is a possibility
of inverse TMR. According to Kim's model, 
transmission probabilities in the parallel ($T_{P}$) and anti-parallel 
($T_{AP}$) magnetic configuration of the two electrodes
(assuming that the spin polarizations of the two electrodes 
are the same) are given as,
\begin{eqnarray*}
T_{P} & = & \frac{2{\gamma}_{+}}{(1+{\gamma}_{+})^{2}}+\frac{2{\gamma}_{-}}{(1+{\gamma}_{-})^{2}}\\
T_{AP} & = & \frac{4\sqrt{{\gamma}_{+}{\gamma}_{-}}}{(1+\sqrt{{\gamma}_{+}{\gamma}_{-}})^{2}}
\end{eqnarray*}
where $\gamma_{+}$ and $\gamma_{-}$ are the transfer rates for majority 
and minority spins respectively.   
When the transmission probability is small, i.e. $\gamma_{\pm}<<1$,
$T_{P}=2(\gamma_{+}+\gamma_{-})$, $T_{AP}=4\sqrt{\gamma_{+}\gamma_{-}}$
which means that the transmission probability in the parallel configuration
is greater than that in the anti-parallel configuration i.e. the TMR is
positive. The conditions for zero TMR or $T_{P}=T_{AP}$ are given as,
${\gamma_{+}}-{\gamma_{-}}=0$
which is a trivial solution and implies that spin polarizations at the Fermi
level for both the electrodes is zero and is applicable for nonmagnetic 
tunnel junctions. The non-trivial solution for zero TMR with spin polarization 
$P\neq0$, resides at the boundary between two regions corresponding 
to $T_{P}>T_{AP}$ and $T_{P}<T_{AP}$ and is given by,
\[({\gamma_{+}}{\gamma_{-}}-1)^{2}-2\sqrt{{\gamma_{+}}{\gamma_{-}}}(1+\gamma_{+})(1+\gamma_{-})=0\]
To be more precise, the combinations $({\gamma_{+}},{\gamma_{-}})$ satisfying
the above equation constitutes a curve in $({\gamma_{+}},{\gamma_{-}})$ space 
with separating the regions corresponding
to normal and inverse TMR. The region close to the origin belongs to 
normal positive TMR. The region away from the origin
contains higher values for ${\gamma_{+}}$ and ${\gamma_{-}}$
which correspond to inverse TMR. 
\begin{figure}
\resizebox{8.5cm}{8cm}
{\includegraphics{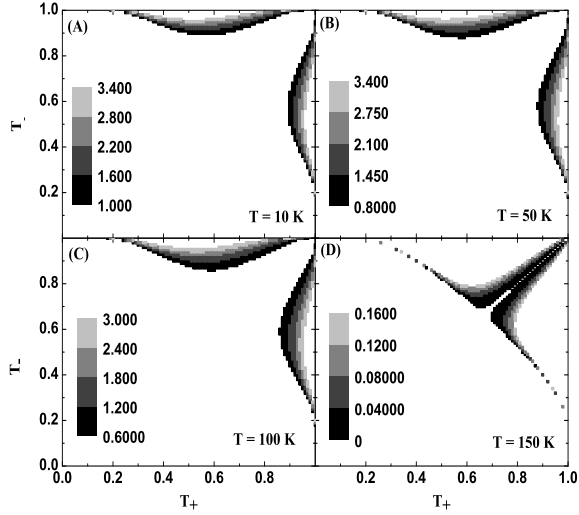}}
\caption{The
allowed values of $T_{\pm}$ which causes inverse TMR for MTJ1 at
$10$ K (A), $50$ K (B), $100$ K (C), $150$ K (D)
respectively within the bias range $\pm 220$ mV.
The corresponding values of inverse TMR in each case is shown by
the shaded map. Up to $100$ K the change in the allowed values of
$T_{\pm}$ is minimal. At $150$ K, there is a sudden congregation of the
phase space near $T_{+}=T_{-}$ line.}\label{fig:phase3}
\end{figure}

\indent
The transmission probabilities for the majority ($T_{+}$) and 
and minority ($T_{-}$) spin band are related to $\gamma_{\pm}$ as follows,
\[T_{\pm}=\frac{4\gamma_{\pm}}{{(1+\gamma_{\pm})}^{2}}\]
Replacing $\gamma_{\pm}$ by $T_{\pm}$ in the expression for
$T_{P}$ and $T_{AP}$, the TMR values [$\Delta R/R_{P}=\{T_{P}-T_{AP}\}/T_{AP}$]
can be calculated numerically for all possible values of $T_{\pm}$.
The transmission probabilities, $T_{P}$ and $T_{AP}$, can be expressed
in terms of $T_{\pm}$ as
\begin{eqnarray*}
T_{P} & = & \frac{1}{2}(T_{+}+T_{-})\\
T_{AP} & = & \frac{{4\sqrt{T_{+}T_{-}}{(1+\sqrt{1-T_{+}})(1+\sqrt{1-T_{-}})}}}{[{\sqrt{T_{+}T_{-}}+{(1+\sqrt{1-T_{+}})(1+\sqrt{1-T_{-}})}}]^2}
\end{eqnarray*}
The theoretically allowed values of ($T_{+}$, $T_{-}$) for
inverse TMR and how the allowed values of ($T_{+}$, $T_{-}$)
evolve with the change in temperature for MTJ1, within the bias range 
$\pm 220$ mV, are shown in Fig:~\ref{fig:phase1} and Fig:~\ref{fig:phase3},
respectively, along with the
corresponding inverse TMR values. Interestingly, there is an upper bound
to the theoretically allowed values of inverse TMR, about $10\%$ which is
a little above the highest experimentally obtained value of inverse
TMR for our system of about $6.5\%$.
Therefore, one can conclude that the sign of magnetoresistance will
decide whether the transport is truly in the ballistic regime 
(transmittivity close to unity).

\indent
Thus, when transmission probability 
is closer to unity i.e. $T_{\pm}\simeq 1$ and there
is an imbalance in the transmission probabilities for the majority spin
and the minority spin, inverse TMR occurs. The contribution due to the parallel
tunneling conduction channel has been neglected for simplicity of
calculation. This will, of course, lead to underestimation of the allowed
values of $T_{\pm}$ particularly in the high temperature region
where the relative contribution of the tunneling conduction channel will
be substantial. The calculation suggests that, larger the imbalance
between $T_{+}$ and $T_{-}$, the greater is the value of
inverse TMR as can be seen from Fig:~\ref{fig:phase1}. 
Up to $100$ K, the allowed values of $T_{\pm}$ stay away from the 
$T_{+}=T_{-}$ line (Fig:~\ref{fig:phase3}A,B,C). 
However, as the temperature is
increased further, the imbalance in the transfer rates of majority
and minority spins diminishes drastically and the allowed values
congregate near $T_{+}=T_{-}$ (Fig:~\ref{fig:phase3}D). 
The increase in bias also reduces the imbalance between the transmittivities
in the two bands as can be seen from the shaded map for each temperature.
The fact that for MTJ2 the contribution due to pinhole conduction is higher 
compared to MTJ1 is consistent with MTJ2 exhibiting higher value of inverse
TMR.

\indent
Here one is forced to ask questions about the origin of minority spin
states since La$_{0.67}$Sr$_{0.33}$MnO$_{3}$ is supposed
to be a ``transport half-metal''. Although La$_{0.67}$Sr$_{0.33}$MnO$_{3}$ 
is generally considered to be having almost full spin polarization,
Andreev reflection experiments have confirmed the existence of minority spin 
states which will be particularly influential in the ballistic limit of 
transport~\cite{minority}. When the point contact Andreev reflection is 
ballistic the calculated value of spin polarization of 
La$_{0.67}$Sr$_{0.33}$MnO$_{3}$ is less compared to the case when the 
contact is diffusive. Since the minority spin states are more localized 
compared to majority spin states, the minority spins do not contribute to
transport spin polarization when the contact is diffusive
and hence gives higher value of spin polarization. 
In this regard, it would be worth mentioning that the authors,
in a previous article, had proposed the existence of minority spin
tunneling states in La$_{0.67}$Sr$_{0.33}$MnO$_{3}$~\cite{soumik}.
If the values of $T_{+}$ and $T_{-}$ are interchanged the TMR
remains the same. However, the physically acceptable situation
is where $T_{+}$ is greater than $T_{-}$, since the minority
spin states are generally regarded as being more localized
compared to the majority spin states.
\begin{figure}
\resizebox {6.5cm}{4.5cm}
{\includegraphics{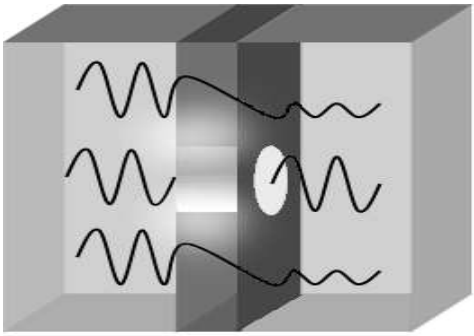}}
\caption{A schematic diagram depicting transport in magnetic tunnel junction
with nanoscale pinhole short. The main two conduction channels are
A: tunneling across the insulating spacer with the wave function decaying
exponentially inside the barrier region resulting in weak transmittivity
and B: transport through pinhole nanocontact with transmittivity
close to unity. At low temperature and low bias voltage the transport
is dominated by current flow through the pinhole where the transmission
probability is close to unity. However, even at the lowest temperature and
bias, the parallel tunneling conduction channel cannot be ignored.
At high temperature and bias, the conduction is dominated by
tunneling across the insulating spacer with weak transmission probability.}
\end{figure}
\indent
The change from inverse TMR to a positive one at higher bias at $150$ K
can be attributed to the fact that at higher
bias electrons tunnel through relatively thin trapezoidal part of
the barrier such that the contribution due to elastic tunneling increases which
gives rise to positive TMR. On the other hand, there are several reasons
for the decrease of inverse TMR at higher bias due to transport through 
the narrow channel.
$1)$ Local generation of heat within the nanocontact region at higher bias
leads to increased thermal spin fluctuation and resistance 
at the nanocontact which reduces the inverse TMR. $2)$ The back-scattering
into the narrow channel increases as a result of larger phonon density
of states at the nanocontact, reducing the transmittivity and hence the
inverse TMR.
In our case, the normal positive TMR is observed at $200$ K, where elastic 
tunneling across the insulating spacer with weak tunneling probability is 
dominant and the electron-phonon interaction pushes the 
transport through the pinhole into diffusive regime. Oxide-based tunnel 
junctions with pinhole shorts are better suited to exhibiting inverse TMR than 
MTJs with transition metal electrodes since in that case there is high 
probability of the pinhole shorts getting oxidized which would lead to weak 
transmittivity through the narrow channel. Another important factor is that 
the mobility in the majority and the
minority spin channel in such systems is vastly different and this 
imbalance is crucial for exhibition of inverse TMR.  

\indent
Lastly, the point which demands serious attention is on the controversy
regarding the existence of ballistic magnetoresistance. 
Even if the claims that the previous reports
of ballistic magnetoresistance suffers from magnetostrictive or magnetostatic
force related artifacts are true there is certainly no reason to conclude
that Ballistic Magnetoresistance does not exist. Kim's model and our
observation should settle the issue that Ballistic Magnetoresistance
can be observed subjected to the following conditions: $1)$ The contact
should be truly ballistic. $2)$ There should be clear imbalance in the
the transmittivities for the majority and minority spin bands. In the
true ballistic limit of transport across ferromagnetic nanocontacts, the 
resistance in parallel magnetic configuration should be higher than the
antiparallel configuration.
\section{Summary}
This article deals with two important aspects of spin dependent transport
in artificial ferromagnetic nanostructures --- firstly, how the existence
of ballistic conduction channels can drastically influence the
magnetoresistive properties of magnetic tunnel junctions, and secondly,
which has more general implications, what should be the sign of 
magnetoresistance
in ferromagnetic nanocontacts in the truly ballistic limit.
We have presented a direct experimental evidence that pinhole
shorts through the insulating spacer in a magnetic tunnel junction
can cause inverse tunnel magnetoresistance when the transmission
probability is close to unity, which is an indicator that Julliere
and Slonczewski models are no longer valid in this regime. 
The relative contributions from the conduction channels due to elastic 
tunneling and ballistic spin conserved transport through the pinholes can 
be changed by proper adjustment of the bias and temperature, which can even 
result in the change of sign of the tunneling magnetoresistance. For practical 
MTJ systems one can always think of the ballistic channel being superimposed
on the tunneling conduction channel causing drastic modification in the 
magnetoresistive response due to the competitive nature of the two
phenomena. 

\indent
This study reconfirms the Andreev reflection results 
concerning the minority spin states influencing transport spin polarization in
La$_{0.67}$Sr$_{0.33}$MnO$_{3}$ and that the minority spin states are
more localized compared to majority spin states.
Another interesting corollary which emerges out is
that for the case of Ballistic Magnetoresistance in ferromagnetic
nanocontacts one can ascertain whether the transport is in the truly
ballistic limit simply looking at the sign of the magnetoresistance.
The results suggest that even if the so called ``Ballistic Magnetoresistance''
might not be due to magnetostrictive or magnetostatic effect one should 
always be careful in determining whether the transport is truly ballistic i.e. 
whether the nanocontact allows for full transmission. On the other hand,
the arguments that there is no ballistic magnetoresistance also seem
to be falsified since Kim's model predicts and our experiments confirm
that ballistic magnetoresistance across ferromagnetic nanocontact is a 
reality.  
\section{Acknowledgements}
The authors would like to acknowledge Mr. S. P. Pai for his technical help
during micro-fabrication of the MTJs.

\end{document}